\documentclass{article}

\usepackage{arxiv}

\usepackage{float}
\usepackage{perpage}
\MakeSorted{figure}
\MakeSorted{table}
\usepackage[usenames, dvipsnames]{color}
\usepackage[ruled,vlined]{algorithm2e}
\usepackage{amsmath}
\usepackage{nomencl}
\usepackage{tcolorbox}
\usepackage{graphicx}
\usepackage{subcaption}
\usepackage{xcolor}
\usepackage{enumitem}
\usepackage{multirow}
\usepackage{array}
\usepackage{textgreek}
\usepackage[nottoc]{tocbibind}

\usepackage[utf8]{inputenc} 
\usepackage[T1]{fontenc}    
\usepackage{hyperref}       
\usepackage{url}            
\usepackage{booktabs}       
\usepackage{amsfonts}       
\usepackage{nicefrac}       
\usepackage{microtype}      
\usepackage{cleveref}       
\usepackage{lipsum}         
\usepackage{graphicx}
\usepackage{doi}

\title{Thermo-mechanical Characterization of Joule Heated Twisted-Coiled Actuators}

\date{}

\newif\ifuniqueAffiliation
\uniqueAffiliationtrue

\ifuniqueAffiliation 
\author{Devang Tavkari \\
	Department of Mechanical and Aerospace Engineering\\
	University of Texas at Arlington\\
	Arlington, TX 76019 \\
	\And
	Paul Davidson \\
	Department of Mechanical and Aerospace Engineering\\
	University of Texas at Arlington\\
	Arlington, TX 76019 \\
	\texttt{paul.davidson@uta.edu} \\
}
\else
\usepackage{authblk}

\newbox{\orcid}\sbox{\orcid}{\includegraphics[scale=0.06]{orcid.pdf}} 
\author[1]{%
	\href{https://orcid.org/0000-0000-0000-0000}{\usebox{\orcid}\hspace{1mm}David S.~Hippocampus\thanks{\texttt{hippo@cs.cranberry-lemon.edu}}}%
}
\author[1,2]{%
	\href{https://orcid.org/0000-0000-0000-0000}{\usebox{\orcid}\hspace{1mm}Elias D.~Striatum\thanks{\texttt{stariate@ee.mount-sheikh.edu}}}%
}
\affil[1]{Department of Computer Science, Cranberry-Lemon University, Pittsburgh, PA 15213}
\affil[2]{Department of Electrical Engineering, Mount-Sheikh University, Santa Narimana, Levand}
\fi

\renewcommand{\shorttitle}{Thermo-mechanical Characterization of Joule Heated Twisted-Coiled Actuators}

\hypersetup{
pdftitle={Thermo-mechanical Characterization of Joule Heated Twisted-Coiled Actuators},
pdfsubject={},
pdfauthor={Devang Tavkari, Paul Davidson},
pdfkeywords={},
}

\begin{document}
\maketitle

\begin{abstract}
In this study, Twisted-Coiled Actuators (TCA) were manufactured by co-coiling a silver coated nylon conductive yarn with a nylon primary mono filament. In the co-coiled TCA, the conductive yarn is used for Joule heating of primary nylon mono filament to provide controlled actuation. A detailed experimental study is conducted to characterize the thermo-mechanical response of TCA, with varying supply of current amplitude and current rates. The results indicates significant rate-dependent hysteritic actuation response to supply current. The details of manufacturing, experiments and results are reported.
\end{abstract}

\section{ Introduction}
In recent years, soft actuators and sensors have seen accelerated development for potential applications in shape-deforming morphing structures, microsensors, and Bio-inspired soft robotics. SMA-driven robotic neck \cite{doi:10.1089/soro.2019.0175,9247113} and PMA-driven soft robots have also been described in the previously reported work\cite{article}. Other studies have explored different types of actuation mechanisms like the electro thermally driven shape memory alloys \cite{MOHDJANI20141078}, carbon nanotube (CNT) composite fiber actuators \cite{doi:10.1126/science.284.5418.1340} \cite{doi:10.1126/science.1222453}, which can actuate relatively heavy loads over a large lengths in a short amount of time. The high cost of manufacturing, low energy efficiency, and hysteretic behavior have led to challenges in implementation of such actuators. Additonally, such actuation mechanisms have limited cycle life and can be hard to control.

A new way to manufacture soft actuators from commonly available high-strength polymer fibers (Fishing line) is discussed by Haines et al. in their seminal work \cite{doi:10.1126/science.1246906}. Twisting and colling high-strength fiber results in a single helix structure, which radially expands and axially contracts on providing an external stimulus of heat. Chemical, electrical, pneumatic, or photothermal inputs can also effectively control these actuators. Elastic rod theory was used to analyze the actuation mechanism in work reported by W. Zheng et al. \cite{article1}. The ability of these fibers to axially contract and radially expand arises due to the polymer fiber's negative thermal expansion coefficient and positive radial expansion coefficient, as explained in \cite{ctx21144815190004911}\cite{article2}. The fiber's molecular structure, wherein entropic forces cause the fiber to contract in length when heated, despite increasing overall volume~\cite{KOBAYASHI1970114}, is the reason behind fiber direction contraction and volumetric expansion. Many highly oriented polymer fibers have negative chain-direction thermal expansion coefficients, including Kevlar and polyethylene~\cite{Rojstaczer1985-md} \cite{Barrera_2005}. Although Kevlar and polyethylene can have very high tensile modulus and strength, their negative thermal expansion coefficients are small. While oriented fibers of nylon 6 and nylon 6,6 do not boast the impressive modulus or strength of Kevlar and polyethylene, their considerably larger negative thermal expansion coefficient can exceed -3.7x$10^{-4} K^{-1}$ at temperatures above their glass transition temperatures \cite{doi:10.1126/science.1246906}, which motivates the implementation of nylon fishing line for manufacturing TCA. 

Twisting nylon fishing line leads to the formation of homochiral coils, heating the fiber then causes the coils to pull together due to the untwist, causing overall contraction in the TCA length. The actuation mechanism has been modeled mathematically, as explained by Lamuta et al. \cite{Lamuta_2018}; however, these calculations are only suitable for materials with isotropic mechanical properties. Fiber directional axial contraction and anisotropic volumetric expansion are prime factors that cause actuation; a fiber volumetrically expands when an external stimulus like joule heat is provided. This expansion causes a significant change in the coiled structure's radius, leading to actuation. 

The primary mechanism of TCA actuation is the volumetric expansion of the primary coil when heated \cite{doi:10.1126/science.1246906}. Of the various means of heating TCAs, the most controllable approach is Joule heating. Joule heat is generated by passing a current through a resistive heating element as an external stimulus to control the actuators. In the present study, a co-coiling method of fabrication is utilized. Thermal input is provided via joule heating of the conductive yarn by passing current through it. Therefore by controlling the current, it is possible to actuate the TCA. However, there are multiple factors that have to be accounted for to achieve accurate actuation control, namely; (a) the influence of current on the temperature of primary filament, (b) the influence of current ramp on the temperate of primary filament, (c) influence of temperature on actuation, and (d) the relationship between current, temperature and actuation. The following sections detail the study to analyze the four factors listed above. Section 2 explains the co-coiling TCA fabrication method. Section 3 details the experimental setup and the experiments conducted. In section 4, the results of the experiments are analyzed.

\section{TCA Fabrication}
There are multiple manufacturing techniques for fabrication of TCA's like twist-induced coiling, mandrel coiling, etc. \cite{9197330}\cite{inproceedings}. Soft actuators are easily fabricated by twist-induced auto-coiling. Fully-coiled structures are then thermally annealed to minimize the strain energy present due to twisting and coiling. TCAs in this work follow the fabrication process described by Haines et al\cite{doi:10.1126/science.1246906}. The difference between prior and current methods lies in the co-coiling of the Joule heating element with the primary filament. A nylon fishing (KastKing Premium Monofilament Fishing Line (20 lb), 0.4mm diameter) line was used as the primary filament, and a silver coated nylon yarn (V-technical textiles 235/36 dtex 4-ply HC+B Conductive yarn) was used as the heating element. Silver-coated nylon yarns are conductive and have high resistivity ($40 \Omega/m \pm 10 \Omega/m$), which does not change during heating. TCAs are manufactured using the following process.

 \begin{figure}[!h]
    \centering
    \includegraphics[width=1\textwidth]{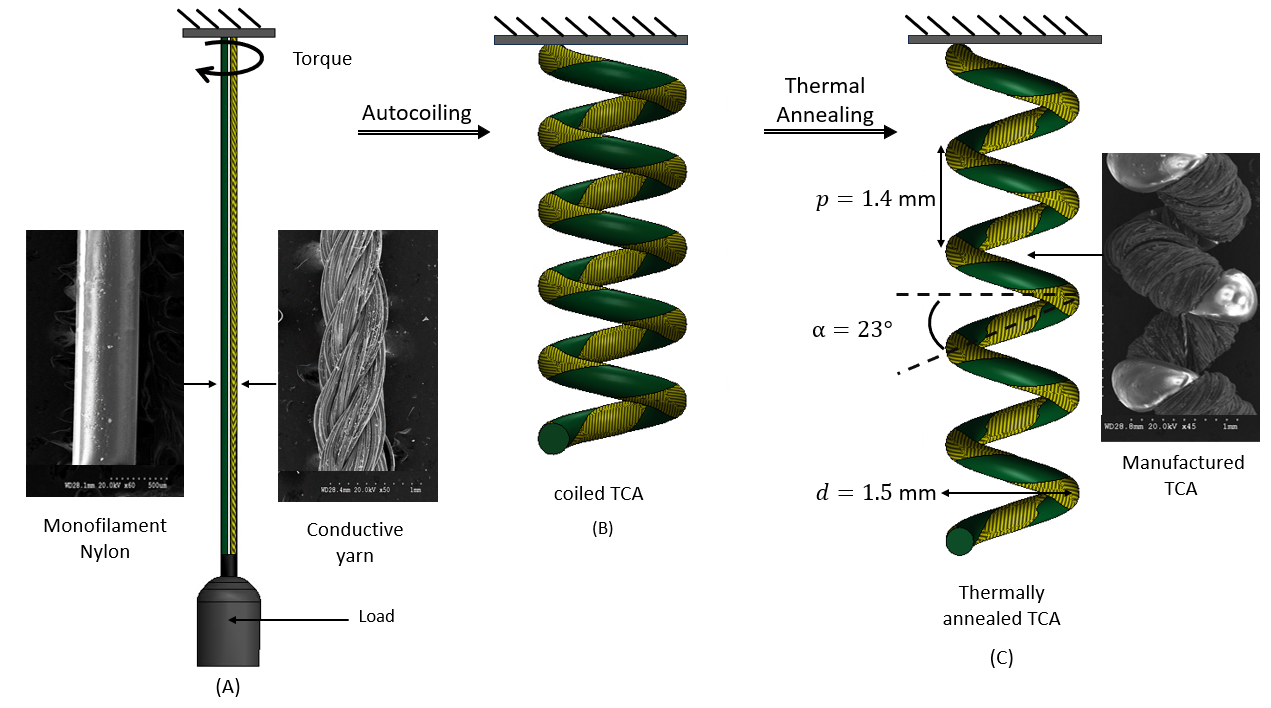}
    \caption{{Fig (A) Straight untwisted monofilament nylon and conductive yarn. Fig (B) Fully coiled TCA .
    Fig(C) Streched and thermally annealed TCA. }}
    \label{fig:Manuf}
\end{figure}

\subsection{Twisting and Coiling}
Twist-induced coiling is a two-step technique for the fabrication of TCAs, which begins by applying torque to equal lengths of nylon fishing line and conductive yarn using a DC motor at one end while applying an axial load and rotational constrain at the other end, as shown in figure \ref{fig:Manuf}A. Load is introduced by hanging weights, with care not to introduce snarls. A 3N end load was applied to straighten the fibers without rupture. Torque is applied via a DC motor with constant 357 rpm. As the filament and yarns are twisted, the filament auto-nucleates a coil, and the yarn wraps around the coil. Figure \ref{fig:Manuf}B shows the Scanning Electron Microscope (SEM) image of co-coiled and auto-nucleated TCA. As can be seen from the image, the conductive yarn wraps around the coiled mono-filament. However, the yarn does not cover the mono-filament completely, with bare spots evenly spaced along the length of TCA. \\
The mechanism of auto-nucleation and complete coiling in mono-filaments of nylon has been extensively studied \cite{PhysRevLett.95.057801}, and for brevity, not repeated here. A TCA's post-coiling geometry depends on the mono-filament material \cite{doi:10.1063/1.4930585}. The mechanical properties of the actuators can be altered by using different mono-filament or by changing the end load. The result of the coiling is a spring-like actuator with an average spring index of 1.53. At this stage, the TCA coil pitch (the distance between two coils) is insufficient for achieving the goal of $30\%$ actuation. The coil pitch of an actuator determines how much a TCA can contract.

 \subsection{Thermal annealing}
 Thermal annealing is the next step which involves stretching and heating the coil to increase the coil pitch and remove any residual stress present in the coil. Thermal annealing begins by mechanically stretching the coiled actuator by 10\% of its overall length and heating it in an oven at $165^\circ C$ for 3 hours and then left to cool at room temperature for 24hrs. Thermal annealing can be repeated multiple times based on the required coil pitch; it is essential to have the heating procedure carried out at accurate temperatures since overheating or underheating can damage the coil structure \cite{BABATOPE19921664}. This process can be repeated to add further stretch to the wire after its initial heat treatment. In this study, TCAs fabricated from 0.4 mm wire were thermally annealed three times to get an actuation of 30\%. Table \ref{tab:experiment1} lists the parameters for TCAs manufactured for this study.

 \begin{table}[]
     \caption{\label{tab:experiment1}Geometric Parameters of TCA.}
              \centering
     \begin{tabular}{lcc}
          \hline
          Geometric parameters&Units & TCA sample\\ 
          \hline
          TCA length (l) & (mm) & 100 \\
          Number of coils (N) &(-) & 70 \\
          Coil Pitch (p) &(mm) & 1.4 \\
          Coil diameter (d)& (mm) & 1.5 \\
          Spring index&(-) & 1.87 \\
          Bias angle ($\alpha$)\ &($^\circ$) & 23$^\circ$ \\
          \hline
     \end{tabular}
 \end{table}
 
\section{Experiment}
\subsection {Experimental Setup}
\begin{figure}[h]
    \centering
    \includegraphics[height = 5cm]{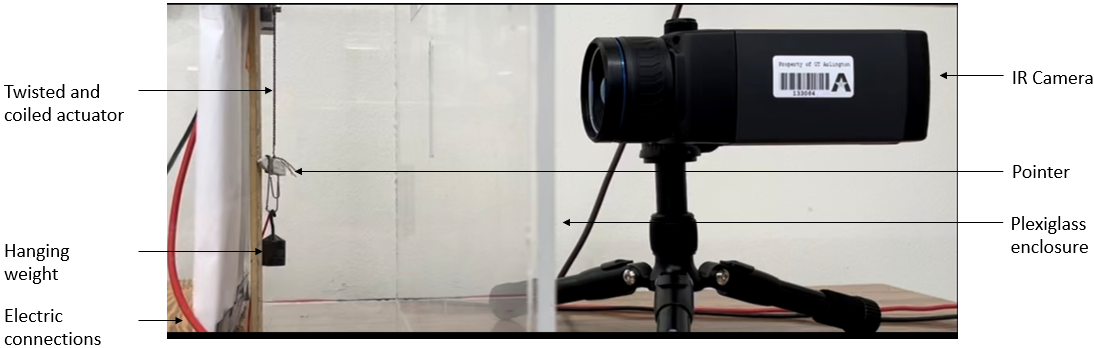}
    \caption{{ The 3D printed testing setup used for performing tests, this setup is placed in a plexiglass enclosure  to avoid heat dissipation during testing.}}
    \label{fig:ExpSetUp}
\end{figure}

 As the actuators are driven thermally, it is necessary to ensure isothermal testing conditions to prevent any influence of ambient airflow on the actuation response. An experiment setup was constructed with a 3D-printed loading frame inside a plexiglass enclosure, as shown in Fig \ref{fig:ExpSetUp}. A thermal camera (high-Resolution Science Grade LWIR Camera FLIR A655sc) was used to measure temperature data, and a camera was used to record the contraction of the actuators due to Joule heating. The camera-recorded contraction data was analyzed using a detection program in Matlab. The current was provided as an input to drive the actuators using a programmable Siglent SPD3303C DC power supply. Tests were conducted with a current ramp ranging from 8.33mA/s to 25mA/s. A small DC current of about 0.1 A to 0.6 A at a voltage of 5V can heat the actuators to about $100^\circ C$ - $120^\circ C $, which was sufficient for actuation. 

\subsection{Study A: Thermal characterization of TCA}
\begin{figure}[]
    \begin{subfigure}{0.4\textwidth}
    \centering
    \includegraphics[width=1.3\textwidth]{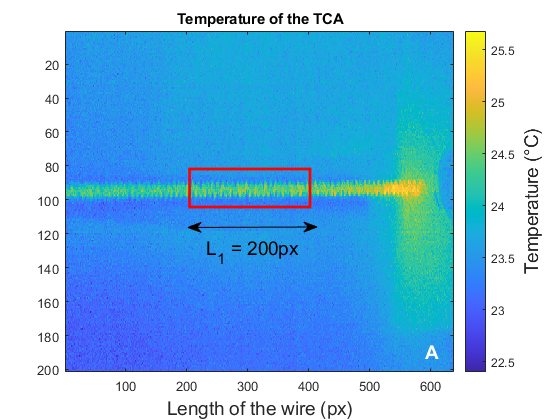}
    \end{subfigure}
    \hspace{1cm}
     \begin{subfigure}{0.4\textwidth}
     \centering
    \includegraphics[width=1.3\textwidth]{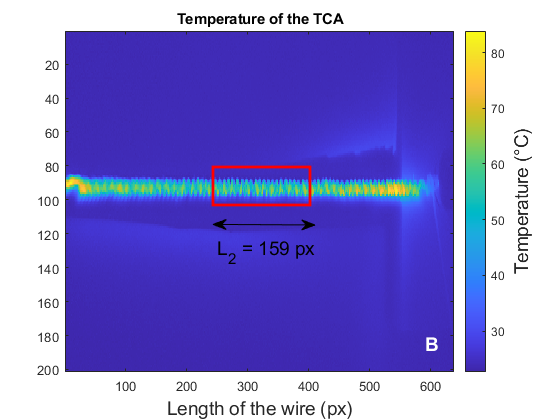}
    \end{subfigure}
    \caption{{Fig(A) Thermal image depicting Dynamic bounding box around the TCA wire used for temperature measurement in the initial frame. Fig(B) Dynamic bounding box around the TCA wire contracting with the TCA wire in the frame with maximum actuation. }}
    \label{fig:ROI}
\end{figure}

The first study was to characterize the temperature profile evolution of the TCA during a fixed current ramp. For this study, the input current was increased at 12.5mA/s up to 0.4A. IR camera captured video frames of the TCA wire every second during the test; the temperature data in the recorded frames were analyzed using a Matlab program. A region of interest (ROI) within each frame was chosen, as shown in Fig \ref{fig:ROI}, and the temperature data were averaged to get the temperature of the TCA. However, as the TCA contracts axially, a new section of TCA enters a static ROI, which causes the temperature average to be higher than the actual temperature of the wire. A dynamic bounding box ROI was used for thermal analysis to ensure temperature was average over the same length of TCA. The dynamic bounding box shrinks in size with the TCA wire to provide an accurate average temperature for a given length of TCA (Fig \ref{fig:ROI}B).

 \subsection{Study B. Electro-mechanical characterization of TCA}
To characterize the influence of current supply and current ramp on actuation, tests were performed where input current was ramped at rates of 8.33mA/s, 12.5mA/s, and 25mA/s, from a minimum of 0A to a maximum of 0.4A, at a constant voltage of 5V. Three replicates of each test were conducted, for which average temperature and total contraction were measured. Table \ref{tab:experiment} shows the experimental test configuration, with a total of 3 experiments and 9 specimens.
  \begin{table}[]
     \centering
     \caption{\label{tab:experiment}Table of experiments conducted.}     
     \begin{tabular}{cccccc}
     \hline
          S.No & Voltage & Current rate & Min. current & Max. current & Replicates  \\
           & V & mA/s & A & A &\\
          \hline
          1 & 5 &8.33 & 0.0 & 0.4 & 3\\
          2 & 5 &12.5 & 0.0 & 0.4 & 3\\
          3 & 5 &25.0 & 0.0 & 0.4 & 3\\
          \hline
     \end{tabular}
 \end{table}
The test sequence is as follows; in the first step, the current supplied to TCA is ramped from minimum to maximum at a given rate. The electrical supply is switched off once the current value reaches the maximum. The experiment is stopped when the average temperature of the TCA returns to room temperature. Throughout the sequence, the temperature and total displacement of the sample were recorded.

\section{Results}
\subsection{Results for study A}
\begin{figure}[!h]
    \centering
    \includegraphics[width=1.05\textwidth]{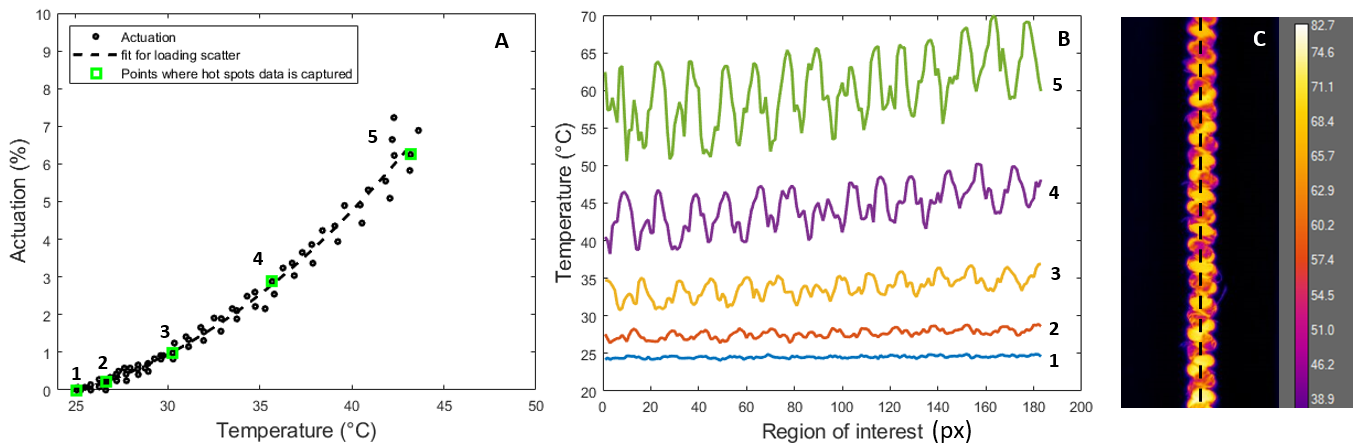}
    \caption{{Fig (A)Plot displaying multiple points at which IR camera recorded hot spot data. Fig (b) Actuation vs Temperature profile for 0.4A test with 12.5 mA/s rate. Fig (c) Thermal image recorded by IR Camera when 0.4A test was performed with 12.5 mA/s rate}}
    \label{fig:studyA}
\end{figure}

The typical TCA thermal profile recorded from the test is shown in figure \ref{fig:studyA}. Figure \ref{fig:studyA}A shows the scatter plot of average temperature vs. percentage actuation of three TCAs (black markers) during loading or actuation. As seen in the figure, as the temperature increases, the actuation also increases. However, the response is non-linear. The non-linear response can be understood by analyzing the temperature along the length of TCA as shown in Fig\ref{fig:studyA}B. The temperature data corresponds to the center line of the thermal profile at various average temperature points represented by green markers in Fig\ref{fig:studyA}A, with point 1 corresponding to the start of the actuation and point 5 to the end of the actuation.

Passing current through the conductive yarn causes heat generation, which is then conducted to the primary filament. As the TCAs were fabricated using the co-coiling method, the Joule heating of the primary filament is uneven, as seen in figure \ref{fig:Manuf}B. This is because conductive yarn, coiled around the primary filament coil, does not completely cover the entire surface. Initially, the temperature difference between the high and low spots within the TCA is small $2^\circ C$. However, as the current increase, the difference also increases by about $15^\circ C$. This indicates that there are two heat conduction mechanisms at play. First is the direct radial conductive heating of the primary filament in contact with the conductive yarn. Second is the axial conductive heating of the primary filament regions not in direct contact with the conductive yarn, as shown in figure \ref{fig:Manuf}C. The two mechanism leads to regions of high (hot spots) and low temperatures within the primary filament. This uneven heating leads to a lag in the expansion of the two regions, thereby causing a lag in actuation response that results in a non-linear response to average temperature. 

Another consequence of hot spots is that even though the average temperature of the TCA can be below the critical glass transition temperature ($110^\circ C-120^\circ C$), the maximum can exceed it. It was experimentally found that current input of 0.1A to 0.4A was enough to actuate the wire to about 45\% of its original length; current values beyond that range would overheat the TCA, which caused the nylon wire to soften. Therefore, the maximum current was limited to 0.4A to avoid failure of TCA due to overheating.

\subsection{Results for study B}
\begin{figure}[!h]
    \begin{subfigure}{0.4\textwidth}
    \centering
    \includegraphics[width=1.3\textwidth]{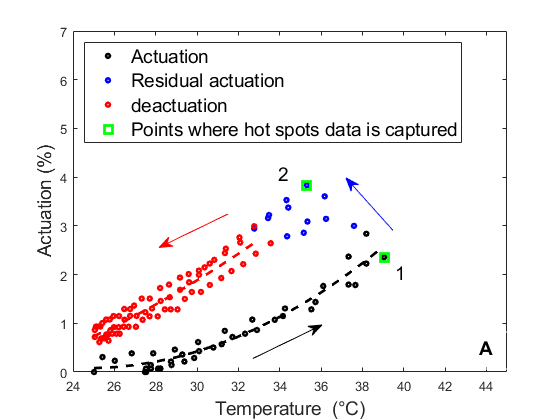} 
    \end{subfigure}
    \hspace{1cm}
     \begin{subfigure}{0.4\textwidth}
     \centering
    \includegraphics[width=1.3\textwidth]{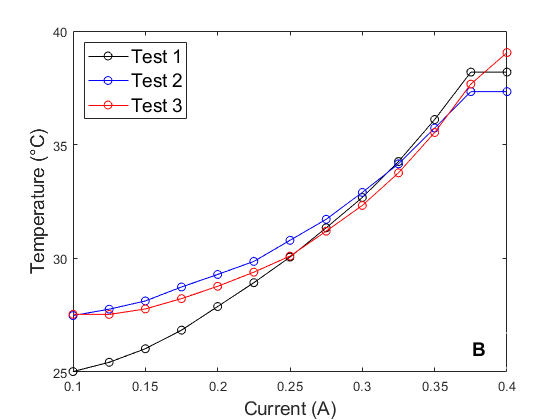}
    \end{subfigure}
    \caption{Fig(A) Actuation vs Temperature scatter plot and fit line showing the actuation, residual actuation and deactuation profile of TCA for 0.4A test with 25mA/s current ramp. Fig (B) Current vs Temperature profile of TCA for 0.4A test with 25mA/s current ramp.}
    \label{fig:StudyB_1}
\end{figure}

\begin{figure}[!h]
    \begin{subfigure}{0.4\textwidth}
    \centering
    \includegraphics[width=1.3\textwidth]{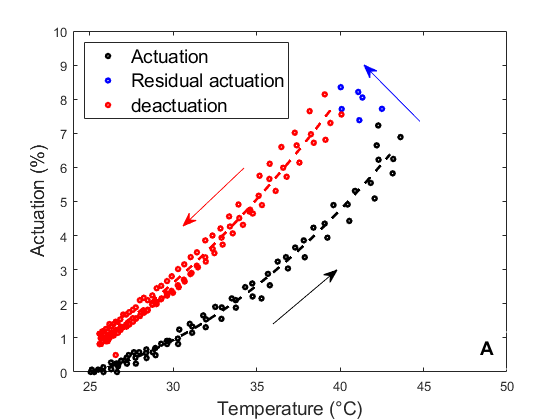} 
    \end{subfigure}
    \hspace{1cm}
     \begin{subfigure}{0.4\textwidth}
     \centering
    \includegraphics[width=1.3\textwidth]{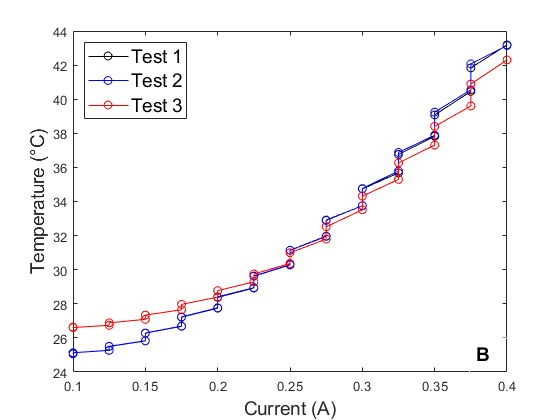}
    \end{subfigure}
    \caption{{Fig(A) Actuation vs Temperature scatter plot and fit line showing the actuation, residual actuation and deactuation profile of TCA for 0.4A test with 12.5 mA/s rate. Fig (B) Current vs Temperature profile of TCA for 0.4A test with 12.5 mA/s rate.}}
    \label{fig:StudyB_2}
\end{figure}

\begin{figure}[!h]
    \begin{subfigure}{0.4\textwidth}
    \centering
    \includegraphics[width=1.3\textwidth]{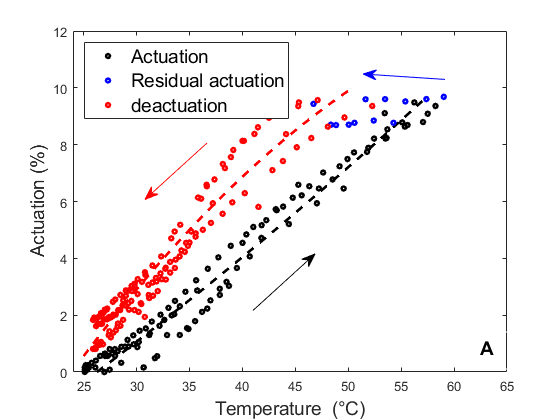}
    \end{subfigure}
    \hspace{1cm}
     \begin{subfigure}{0.4\textwidth}
     \centering
    \includegraphics[width=1.3\textwidth]{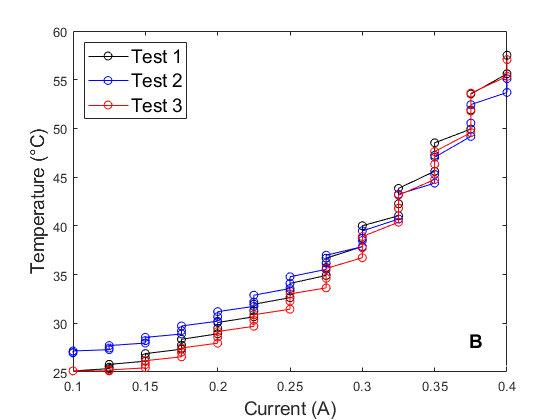}
    \end{subfigure}
    \caption{{Fig(A) Actuation vs Temperature scatter plot and fit line showing the actuation, residual actuation and deactuation profile of TCA for 0.4A test with 8.3 mA/s rate. Fig (B) Current vs Temperature profile of TCA for 0.4A test with 8.3 mA/s rate}}
    \label{fig:StudyB_3}    
\end{figure}

The results of study B are shown in figures \ref{fig:StudyB_1},\ref{fig:StudyB_2},\ref{fig:StudyB_3}. In each figure, sub-figure A shows the contraction versus temperature plot and figure B shows the current versus temperature plot. For figure A, the plot shows the actuation (black), residual actuation (blue) and de-actuation (red) results.

\begin{figure}
    \centering
    \includegraphics[width=0.5\textwidth]{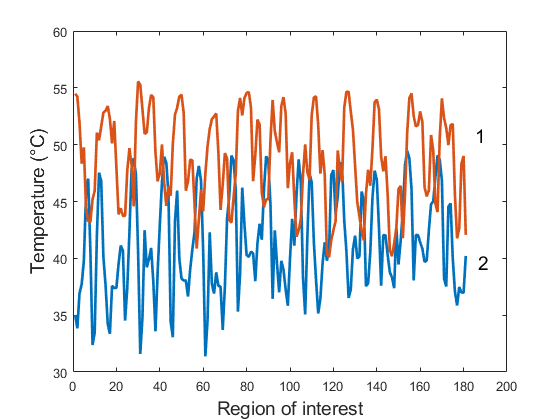}
    \caption{Thermal profile of TCA for points shown in Fig:\ref{fig:StudyB_1}A}
    \label{fig:Studyc_1}
\end{figure}

To understand the results, let us first look at the results for a loading rate of $25mA/s$ as shown in figure \ref{fig:StudyB_1}. The actuation result (black in Fig \ref{fig:StudyB_1}A) is for the case where the current was controlled, as shown in Fig \ref{fig:StudyB_1}B. The supply current varied linearly from 0.1A to 0.4A at 25mA/s. However, the temperature lags and does not increase linearly with the supply current. As in study A, the lag is explained by the hot spots observed during the experiment. Hence, the actuation also does not vary linearly with temperature, as shown in figure \ref{fig:StudyB_1}A. The residual actuation is interesting (figure \ref{fig:StudyB_1}A blue) because even though the current supply is stopped, the heat in hot spots conducts along the primary filament. As the wire thermally equilibrates, the average temperature reduces; however, low-temperature regions in the primary coil increase in temperature, causing an increase in actuation. Fig \ref{fig:Studyc_1} show the linear temperature profile at the maximum actuation and maximum residual actuation, as indicated by the green markers in Fig\ref{fig:StudyB_1}A. As can be observed, the maximum temperature at peak residual actuation point (2) still has a variation of $15^\circ C$ between the max and min temperature, which allows for heat conduction within the coil.  Hence, this phenomenon causes a dynamic lag in temperature with respect to the activation current. After the equilibration of temperatures within the primary coil, the average temperature reduces throughout the coil due to convective heat transfer to the surrounding air. This reverses the contraction and is shown as the red curve in Figure \ref{fig:StudyB_1}A. The lag between current supply, temperature, and actuation is a hysteretic loop.

From the results, it is clear that there is a significant lag between the supplied current and temperature and, in turn, actuation. Additionally, when the same tests are conducted at different supply current rates, the hysteretic response changes (figures \ref{fig:StudyB_2}-\ref{fig:StudyB_3}). As the supply current rate is lowered, the response becomes linear and the loop narrower. This is again explained by the heat transfer mechanism seen in these actuators. With a lower current rate, the primary coil has more time to equilibrate temperatures, which results in near-uniform temperature distribution, as shown in figure \ref{fig:Studyc_1}. Since the temperature throughout the primary coil is now uniform, the actuation is also uniform and linearly varying with temperature increase. The three current ramp rates' results show that the actuation depends on the supply current amplitude and current rate. 
  \begin{figure}[!h]
      \centering
    \begin{subfigure}{0.45\textwidth}
    \includegraphics[width=\textwidth]{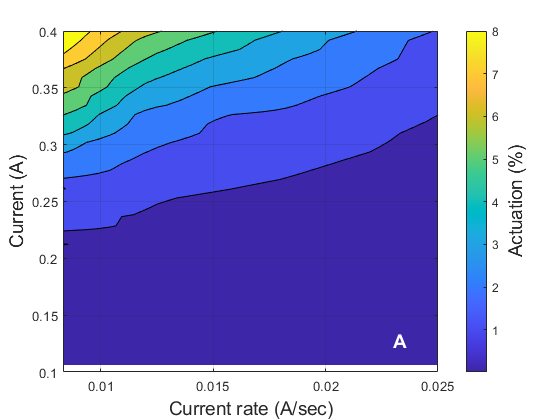}
    \end{subfigure}
     \begin{subfigure}{0.45\textwidth}
    \includegraphics[width=\textwidth]{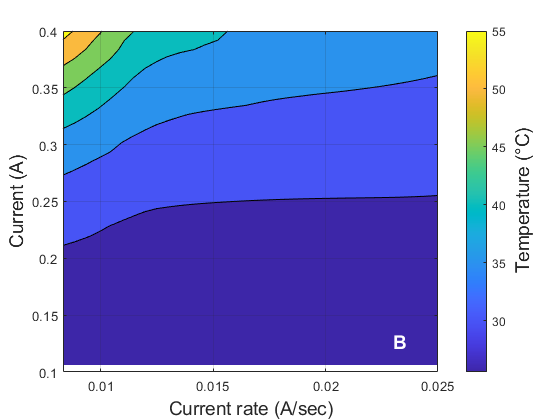}
    \end{subfigure}
    \caption{{ Fig (A) Actuation vs current rate vs current scatter displaying the actuation profile of TCA for 0.4A test. Fig (B) Temperature vs current rate vs current scatter displaying the thermal profile of TCA for 0.4A test.}}
    \label{fig:StudyB_Contour}
\end{figure}
The results of study B can be summarized in figure \ref{fig:StudyB_Contour}, showing the contour plot of temperature response and contraction response with respect to the current and current rate. As seen in the figures, the same level of contraction can be achieved for a lower value of supply current when the current supply rate is also lower. 

\section{Conclusion}
Inexpensive TCA can be manufactured from commonly available polymer fibers using this study's two-stage, co-coiling manufacturing technique. Joule heating using co-coiled silver-coated yarn converts the electrical supply to thermal loads, which in turn actuates the TCAs. The manufacturing model presented in this study can be used to fabricate TCAs with different coil indexes since the heat-treating procedure is the key factor determining the coil index.

To understand the relationship between input current, temperature, and actuation, detailed thermo-mechanical experiments were conducted. The influence of different current rates on heat generation and actuation was also examined. The empirical results indicated a significant rate-dependent hysteresis between the supply current for Joule heating and the actual contraction of TCAs. The loop is narrower at lower supply current rates, and high contraction at lower current is achieved. The reverse is true for higher supply rates. 

With applications of TCAs to soft robotics and mechanism, this study provides an empirical basis for future control algorithms that will require the incorporation of rate-dependent hysteresis. 

\section{Acknowledgements}
The authors would like to thank Prof.Ankur Jain for helpful discussions on thermal measurement techniques and the use of his laboratory for thermal characterization. Author would also like to thank Dr Swapnil Salvi for his assistance during the thermal measurements. This study was supported through PD start-up funds at the University of Texas at Arlington.

\bibliographystyle{unsrtnat}

\begin{thebibliography}{1}
\expandafter\ifx\csname url\endcsname\relax
  \def\url#1{{\tt #1}}\fi
\expandafter\ifx\csname urlprefix\endcsname\relax\def\urlprefix{URL }\fi
\providecommand{\eprint}[2][]{\url{#2}}

\bibitem{doi:10.1089/soro.2019.0175}
Sun J, Tighe B, Liu Y and Zhao J 2021 {\em Soft Robotics\/} {\bf 8} 213--225
  \urlprefix\url{https://doi.org/10.1089/soro.2019.0175}

\bibitem{9247113}
Copaci D, Muñoz J, González I, Monje C~A and Moreno L 2020 {\em IEEE
  Access\/} {\bf 8} 199492--199502

\bibitem{article}
Daerden F and Lefeber D 2002 {\em European Journal of Mechanical and
  Environmental Engineering\/} {\bf 47}

\bibitem{MOHDJANI20141078}
{Mohd Jani} J, Leary M, Subic A and Gibson M~A 2014 {\em Materials \& Design
  (1980-2015)\/} {\bf 56} 1078--1113 ISSN 0261-3069
  \urlprefix\url{https://www.sciencedirect.com/science/article/pii/S0261306913011345}

\bibitem{doi:10.1126/science.284.5418.1340}
Baughman R~H, Cui C, Zakhidov A~A, Iqbal Z, Barisci J~N, Spinks G~M, Wallace
  G~G, Mazzoldi A, Rossi D~D, Rinzler A~G, Jaschinski O, Roth S and Kertesz M
  1999 {\em Science\/} {\bf 284} 1340--1344 (\textit{Preprint}
  \eprint{https://www.science.org/doi/pdf/10.1126/science.284.5418.1340})
  \urlprefix\url{https://www.science.org/doi/abs/10.1126/science.284.5418.1340}

\bibitem{doi:10.1126/science.1222453}
Volder M~F~L~D, Tawfick S~H, Baughman R~H and Hart A~J 2013 {\em Science\/}
  {\bf 339} 535--539 (\textit{Preprint}
  \eprint{https://www.science.org/doi/pdf/10.1126/science.1222453})
  \urlprefix\url{https://www.science.org/doi/abs/10.1126/science.1222453}

\bibitem{doi:10.1126/science.1246906}
Haines C~S, Lima M~D, Li N, Spinks G~M, Foroughi J, Madden J~D~W, Kim S~H, Fang
  S, de~Andrade M~J, Göktepe F, Özer Göktepe, Mirvakili S~M, Naficy S,
  Lepró X, Oh J, Kozlov M~E, Kim S~J, Xu X, Swedlove B~J, Wallace G~G and
  Baughman R~H 2014 {\em Science\/} {\bf 343} 868--872 (\textit{Preprint}
  \eprint{https://www.science.org/doi/pdf/10.1126/science.1246906})
  \urlprefix\url{https://www.science.org/doi/abs/10.1126/science.1246906}

\bibitem{article1}
Wu C and Zheng W 2020 {\em Actuators\/} {\bf 9} 25

\bibitem{ctx21144815190004911}
Choy C~L, Chen F~C and Young K 1981-2 {\em Journal of polymer science.\/} {\bf
  19} ISSN 0098-1273

\bibitem{article2}
Grima J, Zammit V and Gatt R 2006 {\em Xjenza\/} {\bf 11}

\bibitem{KOBAYASHI1970114}
Kobayashi Y and Keller A 1970 {\em Polymer\/} {\bf 11} 114--117 ISSN 0032-3861
  \urlprefix\url{https://www.sciencedirect.com/science/article/pii/0032386170900303}

\bibitem{Rojstaczer1985-md}
Rojstaczer S, Cohn D and Marom G 1985 {\em Journal of Materials Science
  Letters\/} {\bf 4} 1233--1236

\bibitem{Barrera_2005}
Barrera G~D, Bruno J~A~O, Barron T~H~K and Allan N~L 2005 {\em Journal of
  Physics: Condensed Matter\/} {\bf 17} R217--R252
  \urlprefix\url{https://doi.org/10.1088/0953-8984/17/4/r03}

\bibitem{Lamuta_2018}
Lamuta C, Messelot S and Tawfick S 2018 {\em Smart Materials and Structures\/}
  {\bf 27} 055018 \urlprefix\url{https://doi.org/10.1088/1361-665x/aab52b}

\bibitem{9197330}
Tsabedze T, Mullen C, Coulter R, Wade S and Zhang J 2020 Helically wrapped
  supercoiled polymer (hw-scp) artificial muscles: Design, characterization,
  and modeling {\em 2020 IEEE International Conference on Robotics and
  Automation (ICRA)\/} pp 5862--5868

\bibitem{inproceedings}
Mirvakili S, Ravandi A, Hunter I, Haines C, Li N, Foroughi J, Naficy S, Spinks
  G, Baughman R and Madden J 2014 Simple and strong: Twisted silver painted
  nylon artificial muscle actuated by joule heating vol 9056 p 90560I

\bibitem{PhysRevLett.95.057801}
Ghatak A and Mahadevan L 2005 {\em Phys. Rev. Lett.\/} {\bf 95}(5) 057801
  \urlprefix\url{https://link.aps.org/doi/10.1103/PhysRevLett.95.057801}

\bibitem{doi:10.1063/1.4930585}
Yue D, Zhang X, Zhou J and Zhou Y~H 2015 {\em AIP Advances\/} {\bf 5} 097113
  (\textit{Preprint} \eprint{https://doi.org/10.1063/1.4930585})
  \urlprefix\url{https://doi.org/10.1063/1.4930585}

\bibitem{BABATOPE19921664}
Babatope B and Isaac D~H 1992 {\em Polymer\/} {\bf 33} 1664--1668 ISSN
  0032-3861
  \urlprefix\url{https://www.sciencedirect.com/science/article/pii/0032386192910649}

\end{thebibliography}

\end {document}